# Quantum Tunneling in Breather 'Nano-colliders'

V.I. Dubinko [1]

[1] NSC Kharkov Institute of Physics and Technology, Ukraine, vdubinko@hotmail.com

In many crystals with sufficient anharmonicity, a special kind of lattice vibrations, namely, discrete breathers (DBs) can be excited either thermally or by external triggering, in which the amplitude of atomic oscillations greatly exceeds that of harmonic oscillations (phonons). Coherency and persistence of large atomic oscillations in DBs may have drastic effect on quantum tunneling due to *correlation effects* discovered by Schrödinger and Robertson in 1930. These effects have been applied recently to the tunneling problem by Vysotskii et al, who demonstrated a giant increase of sub-barrier transparency during the increase of correlation coefficient at special high-frequency periodic action on quantum system. In the present paper, it is argued that DBs present the most natural and efficient way to produce correlation effects due to time-periodic modulation of the potential well width (or the Coulomb barrier width) and hence to act as breather 'nano-colliders' (BNC) triggering low energy nuclear reactions (LENR) in solids. Tunneling probability for deuterium (D-D) fusion in 'gap DBs' formed in metal deuterides is shown to increase with increasing number of oscillations by ~190 orders of magnitude resulting in the observed LENR rate at extremely low concentrations of DBs. Possible ways of engineering the nuclear active environment based on the present concept are discussed.

## 1. Introduction

The problem of tunneling through the Coulomb potential barrier during the interaction of charged particles is the key to modern nuclear physics, especially in connection with low energy nuclear reactions (LENR) observed in solids [1-4].

The tunneling (a.k.a. transmission) coefficient (TC) first derived by Gamow (1928) for a pure Coulomb barrier is the Gamow factor, given by

$$G \approx \exp\left\{-\frac{2}{\hbar}\int_{r_1}^{r_2} dr\sqrt{2\mu(V(r)-E)}\right\} \quad (1)$$

where $2\pi\hbar$ is the Planck constant, $E$ is the nucleus CM energy, $\mu$ is the reduced mass, $r_1$, $r_2$ are the two classical turning points for the potential barrier, which for the D-D reaction are given simply by $\mu = m_D/2$, $V(r) = e^2/r$. For two D's at room temperature with thermal energies of $E \sim 0.025$ eV, one has $G \sim 10^{-2760}$, which explains a pessimism about LENR and shows the need for some special conditions arising in solids under typical LENR conditions ($D_2O$ electrolysis [1-3], E-cat [4], etc.), which help to overcome the Coulomb potential.

Corrections to the cross section of the fusion due to the screening effect of atomic electrons result in the so-called "screening potential", which acts as an additional energy of collision at the center of mass [5]. The screening potential was measured by the yields of protons or neutrons emitted in the D(d, p)T or D(d, n)$^3$He reactions induced by bombardment of D-implanted solid targets with deuterons accelerated to kinetic energies of several keV, equivalent to heating up to ~$10^7$ K [6]. However, even the maximum screening potentials found in Pt (675 eV), PdO (600 eV) and Pd (310 eV) are far too weak to explain LENR observed at temperatures, which are bellow melting point of solids (E-cat) or boiling point of liquids (electrolysis). Besides, the absence of significant radiation under typical LENR conditions indicates that other reactions should take place, based on interactions between 'slow' particles, which may be qualitatively different from the interactions between accelerated ones.

The most promising and universal mechanism of the stimulation of nuclear reactions running at a low energy is connected with the formation of *coherent correlated states* of interacting particle, which ensures the large probability of the nuclear reactions under conditions, where the ordinary tunnel effect is negligible. These states minimize a more general uncertainty relation (UR) than Heisenberg UR usually considered in quantum mechanics, namely, Schrödinger-Robertson UR [7, 8], which takes into account correlations between coordinate and momentum operators. Correlation effects have been applied to the tunneling problem by Dodonov et al [9] and more recently by Vysotskii et al [10-13] who demonstrated a giant increase of sub-barrier transparency (up to hundreds orders of magnitude!) during the increase of correlation coefficient at special high-frequency periodic action on quantum system.

In this paper, we argue that such an action can be most naturally realized due to time-periodic modulation of the width of potential wells for atoms oscillating in the vicinity of discrete breathers (DBs). DBs are spatially localized large-amplitude vibrational modes in lattices that exhibit strong anharmonicity [14-23]. Due to the crystal anharmonicity, the frequency of atomic oscillations increase or decrease with increasing amplitude so that the DB frequency lies outside the phonon frequency band, which explains the weak DB coupling with phonons and, consequently, their robustness at elevated temperatures. DBs can be excited either

Dubinko V.I.

thermally or by external driving, as was observed experimentally [17, 18] and modelled in various physical systems [19-24].

Studies of DBs in three-dimensional crystals by means of molecular dynamics simulations using realistic interatomic potentials include ionic crystals with NaCl structure [16, 19, 20], diatomic $A_3B$ crystals [21], graphene [22], semiconductors [23] and metals [24, 25]. DBs in biopolymers such as protein clusters have been studied using the coarse-grained nonlinear network model [26].

Presently the interest of researchers has shifted to the study of the *catalytic impact* of DBs on the reaction rates in solids and on the biological functions of biopolymers [26, 27]. Excitation of DBs in solids have been shown to result in a drastic amplification of the reaction rates in their vicinity. Two cases considered up to date include chemical reactions [27-30] and LENR [31]. In the former case, the amplification mechanism is based on modification of the classical Kramers escape rate from a potential well due to a periodic modulation of the well depth (or the reaction barrier height), which is an archetype model for chemical reactions since 1940 [32].

In the latter case, so-called *gap DBs*, which can arise in diatomic crystals such as metal hydrides/deuterides (e.g. palladium deuteride, PdD), have been argued to be the LENR catalyzers [31]. The large mass difference between H or D and the metal atoms provides a gap in phonon spectrum, in which DBs can be excited in the H/D sub-lattice resulting in time-periodic closing of adjacent H/D atoms, which should enhance their fusion probability. The main problem with this mechanism was that unrealistically small separation between atoms (~ 0.01 Å) must be attained in order to increase TC (eq. (1)) by ~ 100 orders of magnitude required for a noticeable LENR rate at the best choice of parameters. Such distances are considerably smaller than the range of conventional chemical forces. However, this estimate did not take into account correlations between coordinate and momentum operators arising in a DB due to cooperative nature of its dynamics. This problem is addressed in the present paper.

The paper is organized as follows. In the next section, formation of coherent correlated states (CCS) under time-periodic action on a particle in the parabolic potential is reviewed [11-13]. In section 3, this model is applied to the evaluation of the TC for the atoms oscillating in the vicinity of DBs and of the corresponding increase of their fusion rate. In section 4, based on the rate theory of DB excitation under $D_2O$ electrolysis and on the modified TC for D-D fusion in the PdD lattice, the fusion energy production rate is evaluated as a function of temperature, electric current and material parameters and compared with experimental data. The results are discussed in section 5 and summarized in section 6.

## 2. Formation of correlated states under time-periodic action on a particle in the parabolic potential

*2.1 Schrödinger-Robertson UR and TC*

The tunneling effect for nuclear particles is closely related to the uncertainty relation (UR), which determines, in fact, the limits of the applicability of the classical and quantum descriptions of the same object. It appears that the well-known and widely used Heisenberg UR is a special case of a more general inequality, discovered independently by Schrödinger [7] and Robertson [8], which can be written in the following form [9]

$$\sigma_x \sigma_p \geq \frac{\hbar^2}{4(1-r^2)}, \qquad (2)$$

$$\sigma_x = \left\langle \left(x - \langle x \rangle\right)^2 \right\rangle, \ \sigma_x = \left\langle \left(p - \langle p \rangle\right)^2 \right\rangle \qquad (3)$$

$$\sigma_{xp} = \langle \hat{x}\hat{p} + \hat{p}\hat{x} \rangle / 2 - \langle x \rangle \langle p \rangle, \qquad (4)$$

where $r = \dfrac{\sigma_{xp}}{\sqrt{\sigma_x \sigma_p}}$, (5)

is the *correlation coefficient* between the coordinate, $x$, and momentum, $p$. At $r = 0$ (non-correlated state) eq. (2) is reduced to the Heisenberg UR, while in a general case, a non-zero $r$ in the UR can be taken into account by the formal substitution

$$\hbar \rightarrow \hbar_{ef} = \frac{\hbar}{\sqrt{1-r^2}}, \qquad (6)$$

which leads to the formal shift of the border between the classical and quantum descriptions of the same object in the transition from non-correlated to correlated state [13].

Then a natural question arises: can nonzero correlations lead to real physical effects? The answer is yes, and the most impressive consequence is a dramatic increase of the tunneling probability, if a true Planck constant in eq. (1) can be replaced by the effective parameter (6). This substitution was justified for a very low barrier transparency (tunneling probability) for the initially uncorrelated state $G_{r=0} \ll 1$ that corresponds to the condition $E \ll V_{\max}$ [13]:

$$G_{r \neq 0} \approx \exp\left\{ -\frac{2}{\hbar_{ef}} \int_{r_1}^{r_2} dr \sqrt{2\mu(V(r) - E)} \right\} \qquad (7)$$
$$= \left(G_{r=0}\right)^{\sqrt{1-r^2}},$$

which is within an order of magnitude close to the result of the exact calculation of the potential barrier transparency using rigorous quantum-mechanical methods [13]. From eq. (7), it follows that when a strongly correlated state with $|r| \rightarrow 1$ is formed, the product of the variances of the particle coordinate and momentum increases indefinitely, and the barrier becomes 'transparent':

$$G_{|r| \rightarrow 1} \rightarrow 1 \text{ even if } E \ll V_{\max}, \qquad (8)$$

Although the substitution $\hbar \rightarrow \hbar_{ef}$ (6) is not quite correct, it clearly demonstrates the high efficiency of using coherent correlated states in solving applied tunneling-related problems in the case of a high potential barrier and a low particle energy $E \ll V_{\max}$, which is typical for LENR.

The physical reason for the huge increase in barrier transparency for a particle in a CCS is the co-phasing of all fluctuations of the momentum for various eigenstates forming the

Dubinko V.I.

superpositional correlated state, which leads to great dispersion of momentum and large fluctuations of kinetic energy of a particle in the potential well.

A CCS can be formed in various quantum systems. The most natural way to form such state is to place a particle in a *non-stationary potential well*. Exact solutions to the non-stationary Schrödinger equation for the particle wave function $\psi(x,t)$ have been found by Vysotskii et al [10-13].

### 2.2 Formation of CCS in non-stationary potential well

A model system considered by Vysotskii et al [11-13] for evaluation of the correlation coefficient is a particle with the mass $M$, coordinate $x(t)$ and momentum $p(t)$ in a non-stationary parabolic potential well (i.e. non-stationary harmonic oscillator),

$$V(x,t) = M(x(t))^2 (\omega(t))^2 / 2, \qquad (9)$$

for which a change of the eigenfrequency $\omega(t)$ was shown to result in an increase of $|r(t)|$. Several scenarios of time evolution $\omega(t)$ have been investigated [11-13], including its monotonic decrease or periodic modulation. The latter regime can be provided, e.g. at a constant well depth $V_{max}$ and the potential well width $L(t)$ that changes periodically resulting in a time-periodic modulation of the eigenfrequency as follows:

$$L(t) = L_0 \left(1 + g_\Omega \cos \Omega t\right), L_0 = \sqrt{8V_{max}/M\omega_0^2} \quad (10)$$

where $L_0$ and $\omega_0$ are the initial parameters of the well before the action of correlated forces, $g_\Omega$ and $\Omega$ are the modulation amplitude and frequency, respectively.

Fig.1 shows that the probability density $|\psi(x,r)|^2$ for the particle localization in the time-periodic well is very narrow for uncorrelated state r=0 (solid black), while it spreads significantly into the sub-barrier region for strongly correlated state r=0.98 at the times of the maximal coordinate dispersion (dash green) [10].

From a detailed analysis [11-13] it follows that the process of formation of strongly correlated coherent state with $|r|_{max} \rightarrow 1$ in response to the action of limited periodic modulation (eq.10) is possible only at any of two conditions: (i) $\Omega = \omega_0$ (resonant formation) or (ii) $\Omega$ is close to $2\omega_0$ (parametric formation): $|\Omega - 2\omega_0| \leq g_\Omega \omega_0$.

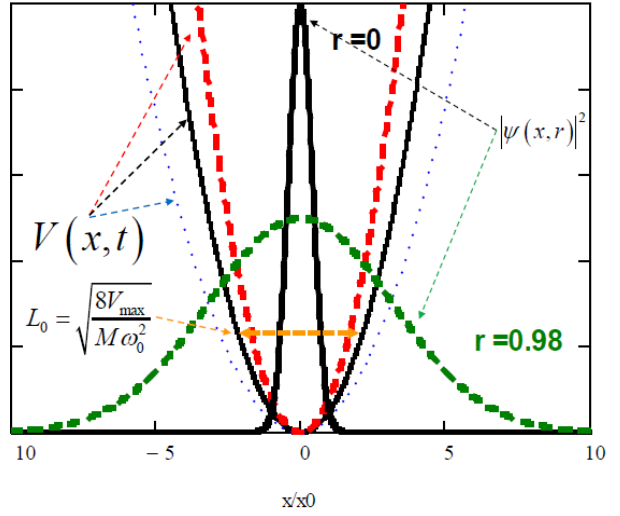

**Fig.1.** Sketch of the time-periodic parabolic potential $V(x,t)$ at different moments of time corresponding to the initial (solid black), minimal (dash red) and maximal (dot blue) well width. Probability density $|\psi(x,r)|^2$ for the particle localization in the well and in the sub-barrier region is shown schematically for uncorrelated state r=0 (solid black) and for strongly correlated state r=0.98 at the times of the maximal coordinate dispersion (dash green) [10]. $x_0 = \sqrt{\hbar/M\omega_0}$ is the half-width of the particle localization in the unperturbed well.

Fig. 2 shows evolution of the correlation coefficient in time under the action of the harmonic perturbation with frequencies $\Omega = \omega_0$ and $\Omega = 2\omega_0$ at $g_\Omega = 0.1$. Correlation coefficient oscillates with time but its amplitude $|r|_{max}$ increases monotonously with the number of modulation cycles, $n = \omega_0 t / 2\pi$, resulting in a giant increase of the tunneling coefficient, as demonstrated in Fig. 3, which shows the TC evaluated by eq. (11) that takes into account both the electron screening [31] and the correlation effects [13]:

$$G^*(L,r) = \exp\left\{-\frac{2\pi e^2}{\hbar_{ef}(r)}\sqrt{\frac{\mu}{2(E + e^2/L)}}\right\}, \quad (11)$$

where L is the minimum equilibrium spacing between D atoms determined by electron screening, E is their kinetic energy (~eV/40 at room temperature) << screening energy ~ $e^2/L$. One can see that the difference in electron screening and the corresponding initial D-D distances in a $D_2$ molecule ($L_0 = 0.74$ Å) and in the PdD crystal ($L_0 = 2.9$ Å) leads to a huge tunneling difference in the initial (uncorrelated) state, in which TC is negligible in both cases. However, with increasing number of modulation cycles, $\hbar_{ef}(r)$ increases according to Fig. 2 resulting in a giant increase of TC up to ~1 in several hundreds of cycles for $\Omega = \omega_0$ and in several dozens of cycles for parametric formation $\Omega \approx 2\omega_0$, which does not require exact coincidence of the frequencies [13].

Dubinko V.I.

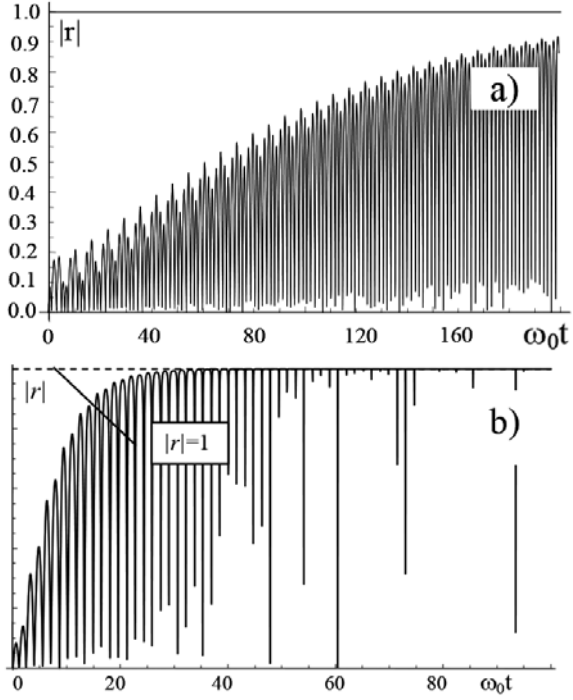

**Fig.2.** Correlation coefficient vs. time of action of the harmonic modulation of the well width given by eq. (10) with a resonant frequency $\Omega = \omega_0$ (a) and parametric resonance frequency $\Omega = 2\omega_0$ (b) at $g_\Omega = 0.1$ [13].

The most important and nontrivial practical question now is how to realize such a periodic action at atomic scale? Modulation of the frequency of the optical phonon modes via excitation of the surface electron plasmons by a terahertz laser suggested in [13] as a driving force for the CCS formation is very questionable [31] (see also discussion in section 5), and it does not explain LENR observed in the absence of the laser driving. In the next section, we will consider a new mechanism based on the large-amplitude time-periodic oscillations of atoms naturally occurring in discrete breathers.

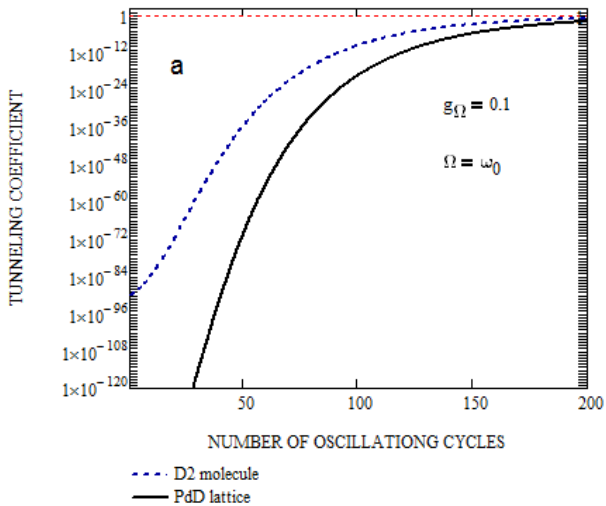

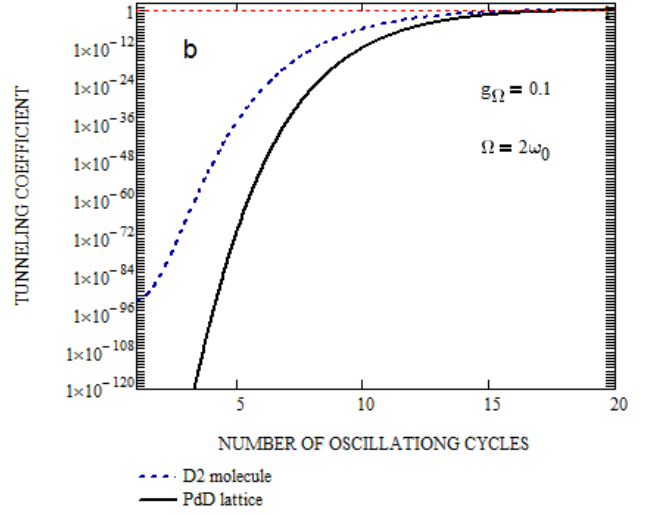

**Fig.3.** Tunneling coefficient increase with increasing number of the well modulation cycles, $n = \omega_0 t/2\pi$, evaluated by eq. (11) for $\Omega = \omega_0$ (a); $\Omega \approx 2\omega_0$ (b), $g_\Omega = 0.1$ for two D-D equilibrium spacings: in a $D_2$ molecule ($L_0 = 0.74$ Å) and in the PdD crystal ($L_0 = a_{PdD}\sqrt{2}/2 \approx 2.9$ Å). $a_{PdD} = 4.052$ Å is the PdD lattice constant at 295 K [33].

## 3. Breather-induced time-periodic action on the potential barrier

In order to develop a mechanism for DB-based LENR in metal hydrides/deuterides (e.g. PdD or NiH) consider their crystal structure in more details. At ambient conditions, Ni/Pd hydrides/deuterides crystallize in FCC structure with the space group of the Rock-salt structure, which is called Fm3m in Hermann–Mauguin notation [33-35] and is shown in Fig.4. The coordination number of each atom in this structure is 6: each heavy (metal) anion is coordinated to 6 light (H/D) cations at the vertices of an octahedron, and vice versa.

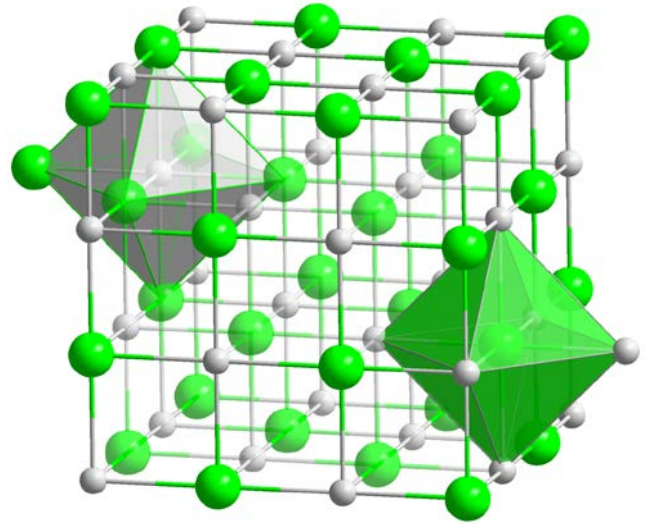

**Fig.4.** The rock-salt or NaCl (halite) structure, in which each of the two atom types forms a separate face-centered cubic lattice, with the two lattices interpenetrating. Heavy (light) atoms are shown by large green (small grey) circles.

Dubinko V.I.

Molecular dynamic (MD) simulations have revealed that diatomic crystals with Morse interatomic interactions typically demonstrate *soft type* of anharmonicity [19], which means that DB's frequency decreases with increasing amplitude, and one can expect to find so-called gap DBs with frequency within the phonon gap of the crystal. The large mass difference between H or D and the metal atoms is expected to provide a wide gap in phonon spectrum (Fig. 5), in which DBs can be excited e.g. by thermal fluctuations at elevated temperatures as demonstrated by Kistanov and Dmitriev [20] for the different weight ratios and temperatures. Density of phonon states (DOS) of the NaCl-type crystal for the weight ratio $m/M = 0.1$ at temperatures ranging from 0 K to 620 K is shown in Fig. 5 (a-d).

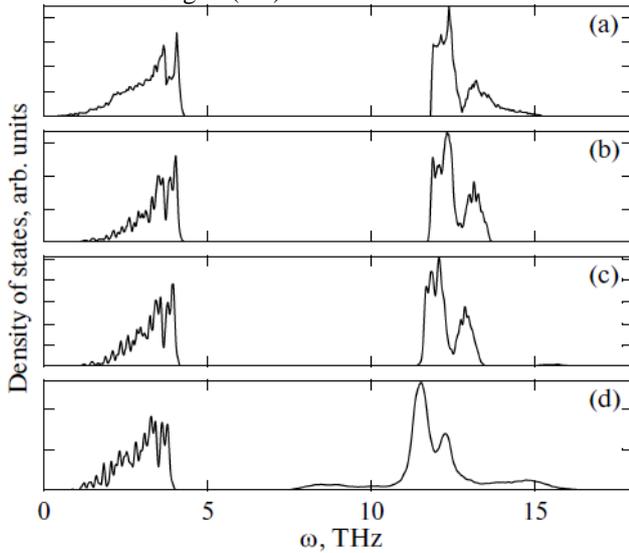

**Fig. 5**. DOS of the NaCl-type crystal for the weight ratio $m/M = 0.1$ at temperatures T = (a) 0, (b) 155, (c) 310, and (d) 620 K. Reproduced from [20] Copyright by APS.

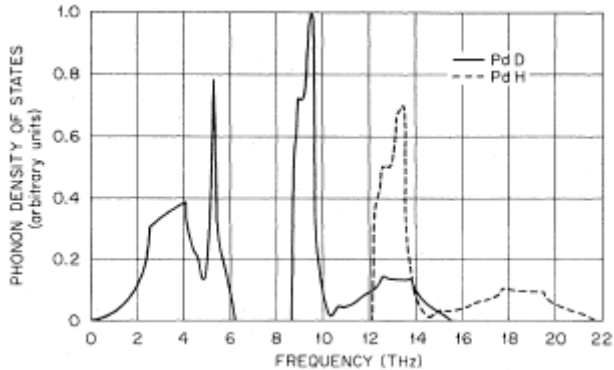

**Fig. 6.** DOS for PdD and PdH crystals based on the force constants obtained from the Born von Karman model [33] fitted to the experimental results for PdD$_{0.63}$ assuming that the forces in PdD and PdD$_{0.63}$ were identical. Reproduced from [33] Copyright by APS.

Fig. 6 shows that DOS for PdD and PdH measured experimentally are qualitatively similar to DOS calculated for the NaCl-type crystal (Fig. 5). First-principles calculations [34] point out that phonon spectra in PdD and PdH are strongly renormalized by anharmonicity.

The appearance of two additional broad peaks in the DOS at elevated temperatures (starting from T = 310 K) can be seen in Fig. 5 (c, d). One of them is in the gap of the phonon spectrum, while another one lies above the phonon spectrum. The appearance of the peak *in the gap* of the phonon spectrum can be associated with the spontaneous excitation of gap DBs at sufficiently high temperatures, when nonlinear terms in the expansion of interatomic forces near the equilibrium atomic sites acquire a noticeable role. As the temperature increases, the lifetime and concentration of gap DBs in the *light atom sub-lattice* increase [21].

The appearance of the peak *above* the phonon spectrum at sufficiently high temperatures can be associated with the excitation of DBs of another type, which manifest the *hard type nonlinearity*. This conclusion agrees with a recent result by Zakharov et al [36] for another type of diatomic crystals, Pt$_3$Al, in which both soft type DBs and hard type DBs were modelled.

Dynamic structure of gap DBs has been revealed in [19] where they have been excited simply by shifting one *light atom* or two neighboring light atoms from their equilibrium positions while all other atoms were initially at their lattice positions and had zero initial velocities. Breather's initial amplitude, $d_0$, was taken from the range $0.01a < d_0 < 0.05a$, where $a = 6.25$ Å is the equilibrium lattice parameter of the model NaCl type structure. In this way, for the weight ratio $m/M = 0.1$, three types of stable DBs have been excited (Fig. 7), frequencies of which are shown in Fig. 8 as the functions of their amplitudes.

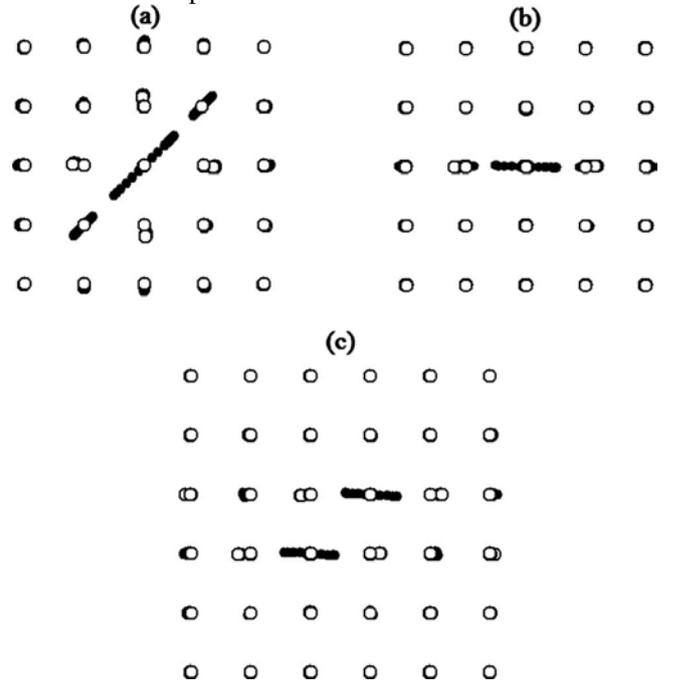

**Fig. 7.** Stroboscopic pictures showing motion of atoms for the DBs of three types: (a) [110]$_1$, (b) [100]$_1$, and (c) [100]$_2$, where figures in brackets describe polarization and the subscript indicates the number of the atoms oscillating with large amplitude. In panels (a) and (b) displacements of the atoms are multiplied by factor 7, and in panel (c) by factor 5. Heavy (light) atoms are shown by open (filled) circles. Reproduced from [19] Copyright by APS.

Dubinko V.I.

Note that all three types of gap DBs are characterized by a high degree of spatial localization of atomic displacements with only one or two atoms having large amplitudes. They shift positions of neighboring atoms, which continue to oscillate with small (harmonic) amplitudes but with frequencies coinciding with DB frequencies shown in Fig 8. Degree of spatial localization of DB increases with decreasing $m/M$ ratio and it remains almost unchanged for about 5000 breather oscillations. During this period DBs radiate their energy very slowly and eventually disappear.

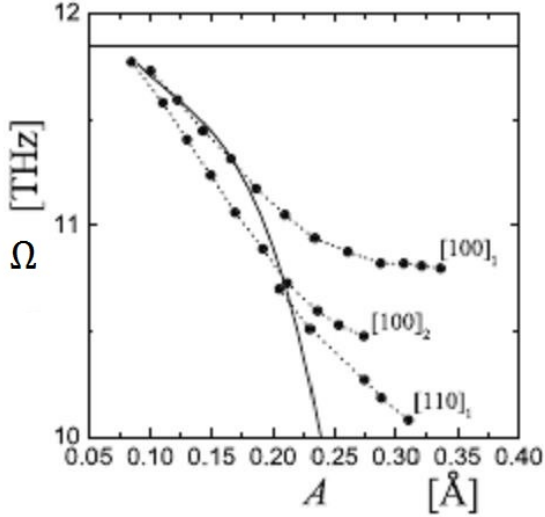

**Fig. 8**. Frequencies, as the functions of the DB amplitudes, A, for the DBs of three types: $[110]_1$, $[100]_1$, and $[100]_2$, excited by simulations [19] for $m/M = 0.1$, where figures in brackets describe polarization and the subscript indicates the number of the atoms oscillating with large amplitude. Solid curve gives $\Omega(A)$ found for the DB $[100]_1$ in the frame of *single degree of freedom model* (see text). Horizontal line gives the upper edge of the phonon gap. Adapted from [19] Copyright by APS.

One can see that the maximum DB amplitude is about 0.34 Å for the polarization $[100]_1$, in which the nearest neighbor (heavy atom) is separated from the light atom by 3.125 Å. The minimum equilibrium spacing between the light atoms along [110] is 4.4 Å, while the maximum DB $[110]_1$ amplitude is about 0.31 Å. An attempt to increase its amplitudes beyond the maximal value lead to its transformation into the DB $[100]_1$, which becomes unstable upon further increase of the amplitude beyond 0.34 Å and decays by radiating phonons [19].

MD modeling in Pt$_3$Al has shown that it is possible to excite hard-type high-frequency DBs by shifting two Al atoms in opposite directions <110>, along which the rows of Al atoms are aligned [36]. These DBs are less localized than gap DBs <100> and they involve several atoms oscillating with large amplitude. What is more, they are highly mobile and can efficiently transfer a concentrated vibrational energy over large distances along close-packed Al directions <110> similar to high frequency mobile DBs in metals [25]. High frequency DBs in NaCl type crystals (manifested by the peak *above* the phonon spectrum in Fig. 5 (d)) have not been modelled so far to our knowledge.

There are two main peculiarities of DBs related to the formation of coherent correlated states, namely, oscillations of atoms comprising a DB are (i) *time-periodic* and (ii) *coherent*, i.e. they have different amplitudes but the same frequency. In the extreme case of DB localized at one light atom, it oscillates with a large amplitude, A in the anharmonic potential well, which determines its frequency $\Omega$ as follows [19]

$$\Omega(A) = \sqrt{\alpha + \frac{3}{4}\beta A^2}, \quad \alpha = \frac{2\gamma_1}{m}, \quad \beta = \frac{2\gamma_2}{m} \quad (12)$$

$$\gamma_1(A) = R_1 A + S_1, \quad \gamma_2(A) = R_2 A + S_2, \quad (13)$$

where $\alpha$ determines the quasi-harmonic eigenfrequency of the potential well, and $\beta$ describes its anharmonicity. Both of them depend on the DB amplitude, since it changes the force constants $\gamma_{1,2}$ of the potential. $\beta$ is positive, which corresponds to hard type of nonlinearity with frequency increasing with A. However, the central atom oscillating with large amplitude shifts positions of neighboring atoms so that $\alpha$ decreases with A resulting in the observed decrease of $\Omega(A)$ shown in Fig. 8. Solid curve shows $\Omega(A)$ for the DB $[100]_1$ in this *single degree of freedom model*, which gives a good fit with 'exact' dashed curve for $A < 0.2$ Å.

Let us apply this model to a DB in the PdD lattice, dispersion curves of which [33] are shown in Fig. 9. In order to understand its physical meaning better, consider dispersion curves of one-dimensional diatomic chain with the force constant $\gamma$ and masses $M > m$ shown in Fig. 10.

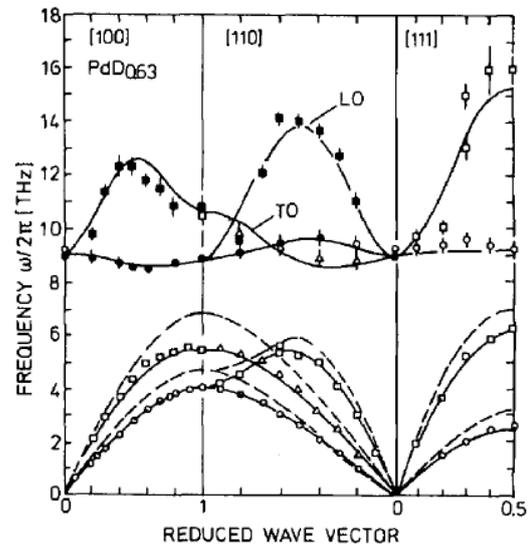

**Fig. 9**. Dispersion curve for PdD$_{0.63}$ measured by coherent neutron inelastic scattering. Solid (open) symbols correspond to phonon peaks measured at 150 K (295 K). LO and TO indicate longitudinal and transverse optical branches. Solid line correspond to the Born von Karman fit [33]. Dashed – acoustic branch of pure Pd. Reproduced from [33] Copyright by APS.

Dubinko V.I.

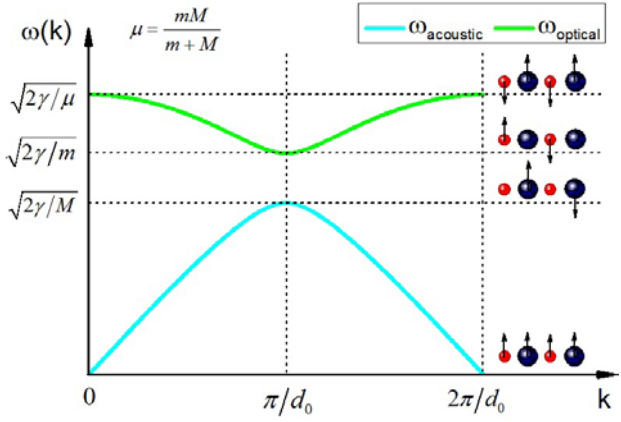

**Fig.10.** Dispersion curve for one-dimensional diatomic chain with the force constant $\gamma$ and masses $M > m$. Arrows show phases of atom oscillations.

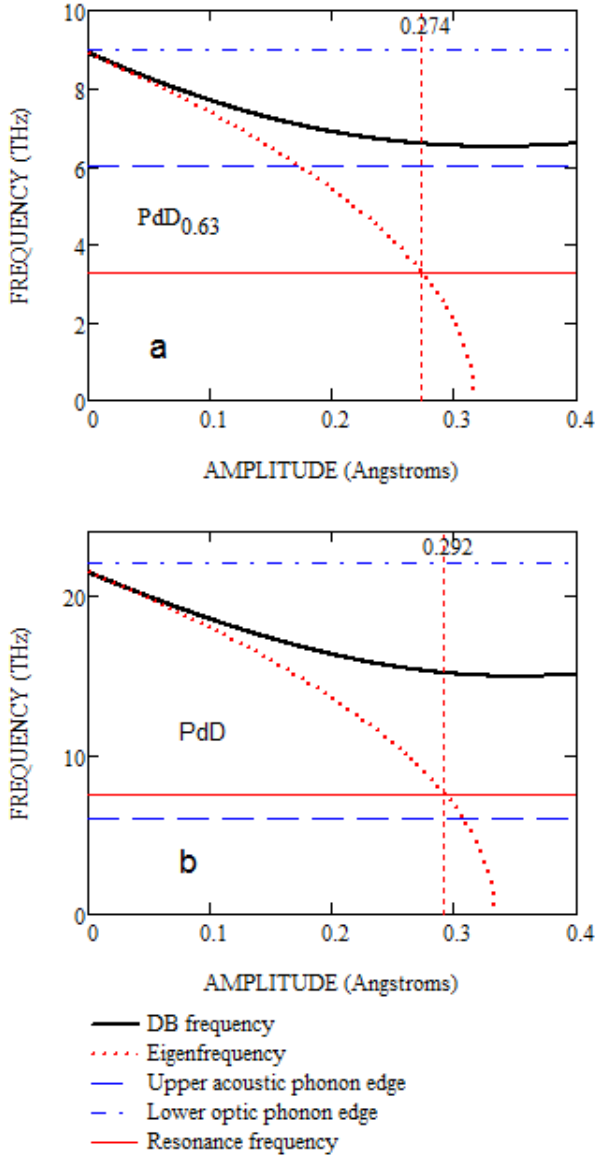

**Fig. 11**. DB frequencies and eigenfrequencies of the potential wells as functions of the DB<110> amplitudes for different force constants assumed for $PdD_{0.63}$ (a) and PdD (b) lattices. <110> is the close-packed D-D direction.

The lower optical phonon frequency (~9 THz) in Fig. 9 is determined by the mass of D, while the higher acoustic phonon frequency (~6 THz) is determined by the mass of Pd and by the corresponding force constants, which are different for D-D, D-Pd and Pd-Pd interactions, in contrast to the one-dimensional model depicted in Fig. 10.

Fig. 11 (a) shows DB<110> frequency $\Omega(A)$ given by eq. (12) and eigenfrequency of (quasi-harmonic) potential wells for neighboring D atoms, $\omega_0(A)$ as functions of the DB amplitude:

$$\omega_0(A) = \sqrt{\frac{2\gamma_1(A)}{m}}, \; \gamma_1(A) = R_1 A + S_1, \qquad (14)$$

evaluated with the force constants assumed to fit dispersion curves of $PdD_{0.63}$ (Fig. 9):

$$R_1 = -0.026 \, \text{eV/Å}^3, \; S_1 = 0.008 \, \text{eV/Å}^2,$$
$$R_2 = -0.017 \, \text{eV/Å}^5, \; S_2 = 0.035 \, \text{eV/Å}^4 \qquad (15)$$

The frequency of the optic modes at the zone center (~9 THz), which determines the maximum DB frequency, is low in comparison with other hydrides, thus implying a weak nearest-neighbor Pd-D force constant in $PdD_{0.63}$ [33]. One of the possible reasons for this broadness of the optic phonon spectrum is due to nonstoichiometry [37].

Fig. 11 (b) shows $\Omega(A)$ and $\omega_0(A)$ evaluated with the force constants fitted to raise the lower optic phonon edge for stoichiometric PdD up to 22 THz and to broaden the phonon gap accordingly:

$$R_1 = -0.143 \, \text{eV/Å}^3, \; S_1 = 0.048 \, \text{eV/Å}^2,$$
$$R_2 = -0.013 \, \text{eV/Å}^5, \; S_2 = 0.143 \, \text{eV/Å}^4 \qquad (16)$$

In the $PdD_{0.63}$ case (Fig. 11(a)), increase of the DB amplitude up to 0.274 Å leads to the excitation in neighboring wells of the harmonic with the frequency~ 3.3 THz equal to half of the main DB frequency (~6.6 THz), which interacts with acoustic phonons below the gap and makes the DB unstable, similar to the NaCl case for *m/M >0.2* [19].

In the PdD case (Fig. 11(b)), increase of the DB amplitude up to the critical value $A_{cr} \approx 0.292$ Å leads to the excitation in neighboring wells of the harmonic with the frequency~ 7.5 THz equal to half of the main DB frequency (~15 THz), which lies above the upper acoustic phonon edge and does not interact with phonons. Such DBs are stable, and they lead to the parametric formation of CCS of deuterons in the neighboring quasi-harmonic potential wells subjected to time-periodic modulation of their eigenfrequencies $\omega_0(A_{cr}) \approx 7.5$ THz by the DB frequency $\Omega(A_{cr}) \approx 15$ THz. As a result of such modulation, D-D fusion is expected to occur in several dozens of DB cycles (Fig. 3b) since the modulation amplitude $g_\Omega \approx A_{cr}/(a_{PdD}\sqrt{2}/2) \approx 0.1$.

Thus, the D-D fusion rate in PdD will be determined by the excitation rate of DBs having amplitudes near the critical value $A_{cr}$, which will be evaluated for typical LENR conditions in the following section.

Dubinko V.I.

## 4. LENR rate under heavy water electrolysis

The DB excitation occurs by thermal fluctuations and by external driving displacing atoms from equilibrium positions. The rate of thermal excitation of DBs having energy $E$ is given by Arrhenius law [27, 31]

$$K_{DB}^{th}(E) = \omega_{DB} k_{DB}^{ef} \exp\left(-\frac{E}{k_B T}\right), \qquad (17)$$

where $k_B$ is the Boltzmann constant, T is the temperature, and $\omega_{DB} \approx \Omega(0)$ is the attempt frequency that should be close to the edge of the phonon band, from which DBs are excited. In the case of gap DBs under consideration, it is about 21 THz (Fig. 11b). $k_{DB}^{ef}$ is the efficiency coefficient for DB excitation.

External driving of the DB excitation can be provided by knocking of surface atoms out of equilibrium position by energetic ions or molecules under non-equilibrium deposition of deuterium under electrolysis. It produces focusing collisions and moving DBs (a.k.a. quodons) that can transfer vibration energy in the crystal bulk [25, 36]. The amplitude of the quasi-periodic energy deviation of atoms along the quodon pathway, $V_{ex}$, can reach almost 1 eV with the excitation time, $\tau_{ex}$, of about 10 oscillation periods, which results in the amplification of DB generation rate proportional to the electric current density $J$ [31]:

$$K_{DB}^{J}(E) = K_{DB}^{th}(E)\left(1 + \left\langle I_0\left(\frac{V_{ex}}{k_b T}\right)\right\rangle \omega_{ex} \tau_{ex}\right) \quad (18)$$

$$\omega_{ex}(F_q) = F_q b^2 \frac{3l_q}{R_P}, \quad F_q = \frac{J}{2e} \qquad (19)$$

where $\omega_{ex}$ is the mean number of excitations per atom per second caused by the flux of quodons, $e$ is the electron charge, $b$ is the atomic spacing. The product $F_q b^2$ is the frequency of the excitations per atom within the layer of a thickness $l_q$ equal to the quodon propagation range, while the ratio $3l_q/R_P$ is the geometrical factor that corresponds to the relative number of atoms within the quodon range in a PdD particle of a radius $R_P$. The coefficient of proportionality between $F_q$ and the electron flux $J/e$ assumes that each electrolytic reaction that involves a pair of electrons, releases a vibrational energy of ~1 eV, which is sufficient for generation of one quodon with energy $V_{ex} < 1$ eV.

Multiplying the DB generation rate (18) by the tunneling probability in a DB, $G^*(L, r)$ (11) and integrating over DB energies one obtains the D-D fusion rate per PdD unit cell:

$$FuR = \frac{1}{\Delta E} \int_{E_{DB}^* - \Delta E}^{E_{DB}^* + \Delta E} K_{DB}^{J}(E) G^*(r) dE, \qquad (20)$$

that dramatically depends on the correlation coefficient, $r$, which, in its turn, strongly depends on the DB amplitude ~ DB energy and the number of DB cycles before decay, $n_{DB}$.

Only a small fraction of DBs can form CCS in their vicinity and act as effective *breather nano-colliders* (BNC). They must have some particular energies, $E_{DB}^* \pm \Delta E$, in order to cause the parametric resonance producing CCS. If $\Delta E \ll k_B T$, eq. (20) is reduced to

$$FuR \approx 2K_{DB}^{J}(E_{DB}^*) G^*(r(E_{DB}^*, n_{DB}))$$

$$\xrightarrow[n_{DB} > n_{DB}^*]{} K_{DB}^{J}(E_{DB}^*), \quad G^*(r(E_{DB}^*, n_{DB}^*)) \approx \frac{1}{2} \qquad (21)$$

where the number of DB cycles required to make the Coulomb barrier 'transparent', $n_{DB}^* \approx 100$ at $g_\Omega = 0.1$.

We consider the following reaction [2]

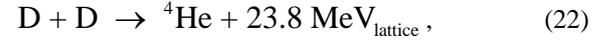

$$D + D \rightarrow {}^4He + 23.8 \text{ MeV}_{lattice}, \qquad (22)$$

which is based on experimentally observed production of excess heat correlated with production of a "nuclear ash", i.e. $^4$He [2, 3]. Multiplying the DB-induced fusion rate (21) by the energy $E_{D-D} = 23.8$ MeV, produced in D-D fusion one obtains the excess energy production rate per atom, $P_{D-D}$ as a function of temperature and electric current:

$$P_{D-D}(T, J) = K_{DB}^{J}(E_{DB}^*, T, J) E_{D-D}, \qquad (23)$$

Usually, the output power density is measured per unite surface of a macroscopic cell, $P_{D-D}^S$, as a function of the electric current density at a fixed temperature and at temperature increasing with $J$, as illustrated in Fig. 12. This is given by the product of $P_{D-D}$, the number of atoms per unit volume, $1/\upsilon_{PdD}$ ( $\upsilon_{PdD}$ being the atomic volume of PdD) and the ratio of the cell volume to the cell surface:

$$P_{D-D}^S(T, J) = P_{D-D}(T, J) \frac{L_S}{\upsilon_{PdD}}, \qquad (24)$$

$L_S$ is the cell size, if cubic, or thickness, in case of a plate.

Fig. 13 shows the LENR output power density DBs as a function of electric current density and temperature evaluated by eq. (24) assuming material parameters listed in Table 1. Comparison of Figs. 13 with experimental data (Fig. 12) shows that the present model describes quantitatively the observed linear dependence of $P_{D-D}^S$ on the current density at a constant temperature as well as the deviation from the linear dependence, if temperature increases with increasing electric current density. Thermally-activated nature of the reactions leading to LENR has been noted for quite a long time [3], and the activation energy was estimated in some cases to be ~0.65 eV. The present model not only explains these observations, but also reveals that the underlying physics is a consequence of the synergy between thermally activated and externally driven mechanisms of the DB excitation in deuterated palladium.

Dubinko V.I.

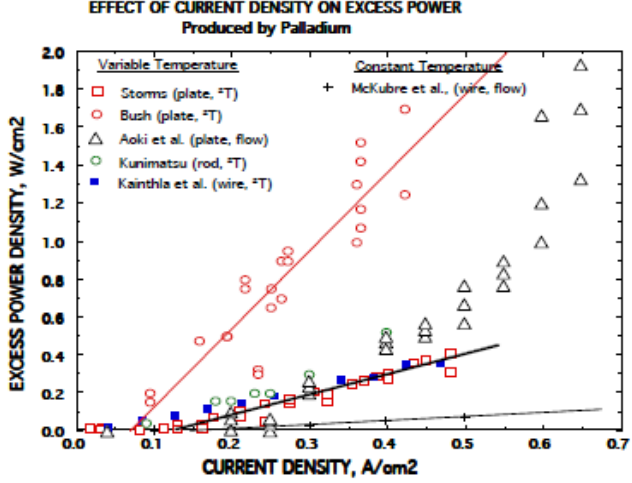

**Fig. 12**. Comparison between several studies showing the effect of current density and temperature on power density [3] Circles, squares and crosses – electrolysis under constant temperature; triangles – $T$ increasing with $J$.

**Table 1.** Material parameters used in the model.

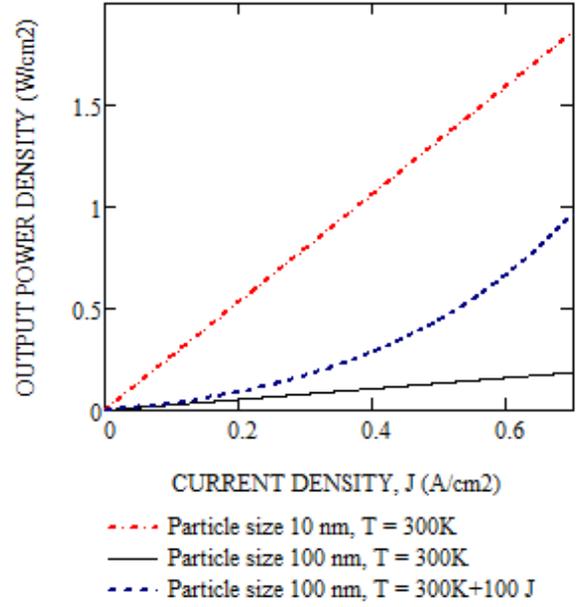

**Fig. 13**. LENR output power density according to eq. (24) as a function of electric current density at constant $T$ and $T$ increasing with $J$ as $T=300K+100J$.

| Parameter | Value |
|---|---|
| D-D equilibrium spacing in PdD, $b$ (Å) | 0.29 |
| DB parametric amplitude, $A_{cr}$ (Å) | 0.292 |
| DB parametric energy, $E_{DB}^*$ (eV) | 1 |
| DB initial frequency, $\Omega(0)$ (THz) | 21 |
| DB parametric frequency, $\Omega(A_{cr})$ (THz) | 15 |
| DB-induced harmonic frequency, $\omega_0(A_{cr})$ (THz) | 7.5 |
| Minimum DB lifetime, $\tau_{DB}^* = n_{DB}^*/\omega_{DB}$ (s) | $6\times 10^{-12}$ |
| DB excitation efficiency, $k_{DB}^{ef}$ | $4\times 10^{-11}$ |
| Quodon excitation energy $V_q \approx V_{ex}$ (eV) | 0.8 |
| Quodon excitation time, $\tau_{ex} = 10/\omega_{DB}$ (s) | $6\times 10^{-13}$ |
| Quodon propagation range, $l_q = 10b$ (nm) | 2.9 |
| Cathod size/thickness (mm) | 5 |

Dubinko V.I.

## 5. Discussion

The main message of this paper is that DBs present the most efficient way to produce CCS due to time-periodic modulation of the potential well width (or the Coulomb barrier width) and hence to act as BNC triggering LENR in solids. The BNC concept proposed in a previous work [31] did not take into account correlation effects, and hence, unrealistically small separation between atoms (~ 0.01 Å) would have to be attained in order to enhance the LENR rate up to a noticeable level. Fig. 14 demonstrates effect of CCS in the BNC model manifested by a number of DB cycles required to produce experimentally observed LENR rate ~1 W/cm$^2$. It can be seen that in the modified model, the DB lifetime plays much more important role than the tunneling D-D spacing, and that DB amplitude of several fractions of angstrom is sufficient to produce required effect, if CCS parametric conditions are met.

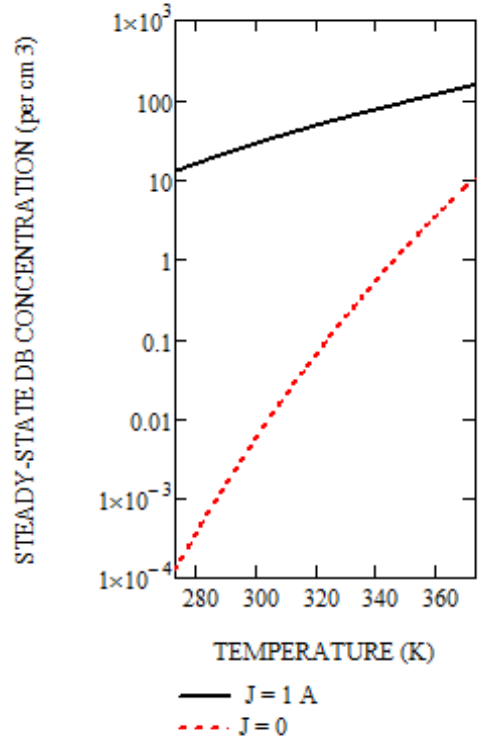

Fig. 15. DB steady-state concentration vs. temperature at zero and typical LENR electric current.

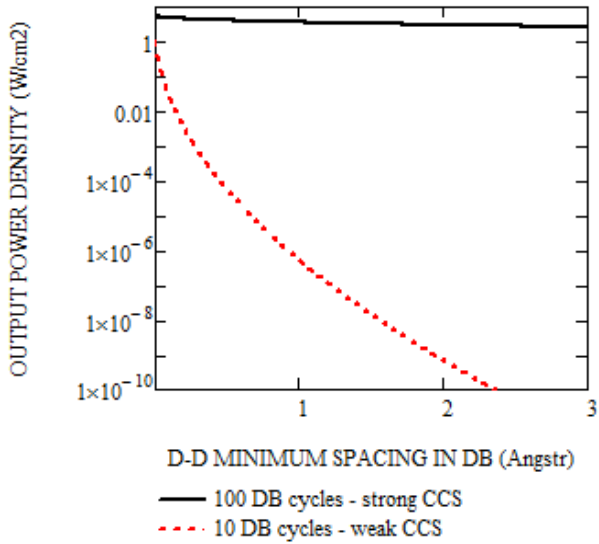

**Fig. 14**. LENR output power density according to eq. (24) as a function of the tunneling D-D spacing at strong and weak CCS.

The correlation effects can make DBs so efficient that their steady-state concentration required to produce observed LENR rate, $K_{DB}^{J}\left(E_{DB}^{*}\right)\tau_{DB}^{*}/\upsilon_{PdD}$, can be extremely low (Fig. 15).

Medvedev et al [38] has demonstrated by means of MD simulations that gap DBs can be excited in the Al sub-lattice of Pt$_3$Al under the action of *time-periodic external driving*. Time-periodic shaking of the surface atoms at frequencies near the optic phonon edge resulted in the DB excitation in the sub-surface layers. These findings point out at the possibility of LENR stimulation by external time-periodic excitation of surface atoms. That is what has been actually realized in "Terahertz" laser experiments [39] on the stimulation of nuclear reaction at a joint action of two low-power laser beams with variable beat frequency ranging from 3 to 24THz on the cathode surface during the D$_2$O electrolysis in the PdD system.

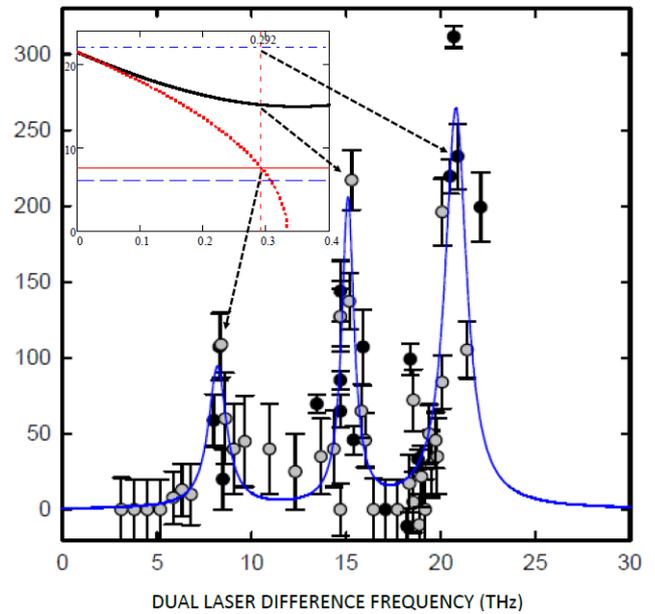

**Fig. 16.** Excess power (mW) under joint action of two low-power laser beams with variable beat frequency on the surface of the Pd cathode during the electrolysis in heavy water [39]. The inset (from Fig. 11b) shows parity between critical DB-induced frequencies and the resonance frequencies [39] designated by dashed arrows.

Dubinko V.I.

Fig. 16 shows the experimental frequency dependencies of the excess power in these experiments. Three main resonances of excess energy released at ~ $8\pm1$ THz, $15\pm1$ THz and $21\pm1$ THz correlate with the DB-induced harmonic frequency, $\omega_0(A_{cr}) \approx 7.5$ THz, DB parametric frequency $\Omega(A_{cr}) \approx 15$ THz and DB initial frequency, 21 THz, respectively (see the inset in the figure). According to the present model, the highest resonance is the biggest, since it is caused by amplification of DB excitation at the edge of optic phonon band. The medium resonance is due to tuning action of external driving on the DB frequencies: it increases the fraction of DBs with parametric frequency. The lowest resonance is due to tuning of harmonic frequencies by external driving: it increases the fraction of D atoms subjected to the parametric action by DBs.

The atoms are shaken by laser beams via excitation of the surface electron plasmons as suggested in [39]. It explains the necessity of external magnetic field for producing resonance effects [39, 13]. However, the direct modulation of the frequency of the optical phonon modes by plasmons proposed in [13] as a driving force for the CCS formation is very questionable [31], and it does not explain LENR observed in the absence of the laser driving at slightly higher electric current or temperature [39]. In the present model, the laser driving acts just as a tuning tool for the CCS formation by DBs induced by temperature and electric current.

One of the new and important consequences of the present model is that it offers a principal explanation for the critical D loading required for LENR. It is known that excess heat usually does not appear until the loading is near 0.83 [39], whereas neuron scattering measurements of the phonon spectra have been done for non-stoichiometric $PdD_{0.63}$, the DOS of which (Fig. 9) does not support formation of CCS by the present mechanism, since its phonon gap is not sufficiently broad (Fig. 11a). Our hypothesis is that mechanical stresses arising in $PdD_x$ above the critical loading x > 0.83 can make the phonon band similar to that shown in Fig. 11b thus switching on the DB-induced formation of CCS.

Structural information on ball milled magnesium hydride from vibrational spectroscopy and ab-initio calculations [40] has shown that the high-energy part of the vibrational spectrum is rather sensitive to stresses induced by for instance ball milling. The structure of PdD at extreme loading is similar to that after the ball milling. As noted by Storms [3], "cracks and small particles are the Yin and Yang of the cold fusion environment", which seems to support the present hypothesis.

Another factor concerning the role of the crystal disorder in LENR is a striking *site selectiveness* of DB formation in the presence of spatial disorder [26, 30, 41]. It means that the process of loading or special 'nano-treatment' creates the disordered cluster structures, which may be enriched with sites of *zero or small threshold energies* for the DB excitation. Such sites are expected to become the *nuclear active cites*, according to the present model. The most important consequence of this hypothesis is that it may offer the ways of *engineering* the nuclear active environment based on the MD modeling of DB creation in nanoparticles and disordered structures.

## 6. Conclusions and outlook

Persistent spatially localized vibrations of nonlinear origin known as discrete breathers (DBs) that can be excited generically in many-body nonlinear systems are proposed to to produce coherent correlation states (CCS) due to time-periodic modulation of the potential well width and hence to act as breather 'nano-colliders' (BNC) triggering low energy nuclear reactions in solids. In particular, tunneling probability for deuterium (D-D) fusion in 'gap DBs' formed in metal deuterides has been shown to increase with increasing number of oscillations by ~190 orders of magnitude resulting in the observed LENR rate at extremely low concentrations of DBs.

The present model describes the observed linear dependence of the excess power output on the current density under heavy water electrolysis at a constant temperature as well as its exponential increase with increasing temperature, which can be the basic LENR mechanism in the hot CAT-type installations.

The proposed mechanism of CCS formation near the gap DBs requires sufficiently broad phonon gap that is not observed below the critical D loading ~0.83 examined so far. Further investigations of DOS and DBs in the extreme conditions of LENR are required.

An alternative mechanism of the DB-induced CCS formation may involve high frequency (hard type) DBs, manifested by the peak *above* the phonon spectrum in NaCl type crystals. Atomistic modeling of DBs of various types in metal hydrides/deuterides is an important outstanding problem since it may offer the ways of *engineering* the nuclear active environment.

### Acknowledgements
The author is grateful to Sergey Dmitriev, Vladimir Vysotskii and Mikhail Bogdan for helpful discussions and valuable criticism.